# Event Extraction Based on Deep Learning in Food Hazard Arabic Texts

Fouzi Harrag
Computer Sciences Department,
College of Sciences,
Ferhat Abbas University,
Setif, Algeria,
fouzi.harrag@univ-setif.dz

Selmene Gueliani
Computer Sciences Department,
College of Sciences,
Ferhat Abbas University,
Setif, Algeria,
selmengueliani@gmail.com

*Abstract*— **Social Media websites have disseminated digital devices to the public, making information sharing easier and faster. Exchanging textual data is the most popular communication among social media users. It has become a necessity for treatment. Event extraction on the other hand indicates an understanding of events across social media posts streams. Event extraction helps to take faster corrective action in natural disasters, and may save lives. The main objective of the task is to develop specific model to detect and extract the events (incidents) identified in the digital text. We proposed here a model based on deep recurrent networks to extract the events from social media feeds.**

## INTRODUCTION

Since a long time ago, people suffer from emerging food hazards. This kind of risks is the main problem that threatens people's health and in the majority of cases lead to the risk of death. Emerging food hazards are problems that occurred in the past or newly appeared among population and quickly spreading in a different geographic range through bones in fish, flaking paint, hair, dirt, metal fragments and nails. The majority of these illnesses result from biological hazards such as bacteria, viruses and parasites. Consequently, food hazards are increasing due to the problem of air pollution, climate changes around the human environment, lack of awareness among community members and lack of resources for informing peoples about the spread of these threats.

### A. Problematic and motivation

Social media is considered as an important source of information. The amount of information published every day on social media is highly increasing. The investment of this mine of data has become a necessity, where the good mining will give insights about the future. Food hazard in the other side is a big threat to human health safety, we aim in this study to use the textual data shared in social media website to extract events related to food hazards. The purpose of this study will be to propose an information extraction based system using recurrent neural networks in event extraction from social media feeds.

### B. Project goals

This project aims to deliver a system that be able to extracts events that has relation with food hazards using social media networks, which are currently considered as the significant resource of information around the world. This work will be useful in the future to identify food hazards events that are propagated over social media networks and news To track all information related to this event and its affected elements. This study will also help the health's experts and authorities to prevent health catastrophes caused by food contamination, by enhancing their level of knowledge and information about food hazards and contamination by determining the symptoms epidemic geo-locations.

The paper is organized as follow: Section 2 gives a general overview about food hazard field and data mining techniques used for detecting events from social media. Section 3 is dedicated to related works. Section 4 defines the proposed system and architectures for problem solving. Section 5 concludes the papers.

## BACKGROUND

This Section describes different concepts such as food hazard, social media, data mining, text mining, information extraction and recurrent neural networks.

### A. Food Hazard

Food is any substance [1] consumed to provide nutritional support for the body. It is usually of plant or animal origin, and contains essential nutrients, such as fats, proteins, vitamins, or minerals. The substance is ingested by an organism and assimilated by the organism's cells to provide energy, maintain life, or stimulate growth. Historically, people secured food through two methods: hunting, gathering, and agriculture. Today, the majority of the food energy required by the ever-increasing population of the world is supplied by the food industry. Food safety and food security are monitored by agencies like the International Association for Food Protection, World Resources Institute, World Food Program, Food and Agriculture Organization, and International Food Information Council. They address issues such as sustainability, biological diversity, climate change, nutritional economics, population growth, water supply, and access to food. The right to food is a human right derived from the International Covenant on Economic, Social and Cultural Rights (ICESCR), recognizing the "right to an adequate standard of living, including adequate food", as well as the "fundamental right to be free from hunger". Food hazards can be classified into three main types: physical, chemical and biological hazards [2].

- **Physical hazards:**

Physical hazards is any foreign matter unintentionally introduced to food or a naturally occurring object that could cause illness or injury to the person consuming the food item. Examples of physical hazards could include bones in fish, flaking paint, hair, dirt, metal fragments and nails. Sources for contaminants include raw materials, badly maintained facilities and equipment, improper production procedures and poor employee practices [2].

- **Chemical hazards:**

Natural and manufactured chemicals can cause people to become sick if they have contaminated food at the source or during processing. Chemical hazards can be divided into two categories: chemical agents and toxic metals. Examples of toxic metals include copper, zinc used in galvanized containers, and tin used in pewter [2].

- **Biological hazards:**

While physical and chemical hazards have potential to cause foodborne illness, the majority of these illnesses result from biological hazards such as bacteria, viruses and parasites (referred to collectively as pathogens) [2].

Foodborne illness, commonly called "food poisoning", is caused by bacteria, toxins, viruses, parasites, and prions. Roughly 7 million people die of food poisoning each year, with about 10 times as many suffering from a non-fatal version. The two most common factors leading to cases of bacterial foodborne illness are cross-contamination of ready-to-eat food from other uncooked foods and improper temperature control. Less commonly, acute adverse reactions can also occur if chemical contamination of food occurs, for example from improper storage, or use of non-food grade soaps and disinfectants.

In more recent years, a greater understanding of the causes of food-borne illnesses has led to the development of more systematic approaches such as the *Hazard Analysis and Critical Control Points (HACCP)*, which can identify and eliminate many risks [3].

- **Hazard Analysis and Critical Control Points (HACCP)**

HACCP is a management system designed to address food safety through the analysis of physical, chemical, and biological hazards. HACCP focuses on prevention rather than the inspection of final products and involves identifying possible hazards that can cause injury or illness, monitoring these hazards, and implementing corrective action if deviations have occurred.

HACCP examines all phases of the food system from raw material production, procurement and handling, to manufacturing, distribution, and consumption of the finished product. Thus, food hazard events detection becomes a most prominent role in the food safety environment. The detection of food hazard is a most complex issue, which is analyzed and evaluated by the various researchers using different methodologies.

*A. Data mining techniques for event detection from social media*

Social media and internet-based data are becoming a main source for the agencies and people who are looking for information about health and diseases. Social media is the best practical solution for managing and pursuing food hazards. Social media plays the main role in actual time to allow public health systems to do their functions in monitoring and preventing food hazards. To invest these precious mines of data, we need to exploit artificial intelligence and data mining techniques.

- **Data Mining**

We are actually living in the age of data where electronic devices around us record our information, our decision, choices in the supermarket, our financial habits and many more information. With this massive amount of data, we need to extract information or more specifically knowledge from it that can help us to make decisions. This necessity has led to the birth of an approach known as Data Mining. This new approach was the fruits of work of many years in different research fields such as Artificial Intelligence, Machine Learning, Data Analysis and Databases systems.

- **Deep Learning**

Deep learning, as a new area of machine learning research, is a process that allows the computer to learn to perform tasks in a very similar way to the human brain. Currently, deep learning (DL) methods have had a profound impact on computer vision and image analysis applications, such as image classification, segmentation, image completion and so on.

1. **RNN &LSTM (Long Short-Term Memory)**

Recurrent Neural Networks are used when any model needs context to be able to provide the output based on the input. Sometimes the context is the single most important thing for the model to predict the most appropriate output. Long Short-Term Memory is an artificial recurrent neural network architecture used in the field of deep learning. Unlike standard feed-forward neural networks, LSTM has feedback connections. It cannot only process single data points (such as images), but also entire sequences of data (such as speech or video). For example, LSTM is applicable to tasks such as un-segmented, connected handwriting recognition or speech recognition.

2. **CRF (Conditional Random Fields)**

Conditional Random Fields are a discriminative model, used for predicting sequences. They use contextual information from previous labels, thus increasing the amount of information the model has to make a good prediction. [4]

3. **Word Embedding**

Word Embedding is one of the most popular representation of document vocabulary. It is capable of capturing context of a word in a document, semantic and syntactic arity, relation with other words, etc. more detai word embedding's are Collective term for models that learned to map a set of words or phrases in a vocabulary to vectors of numerical values. Neural Networks are designed to learn from numerical data [5]. Most popular off-the-shelf word embedding models in use today are: Word2Vec (by Google), Glove (by Stanford), and fastText (by Facebook).

4. **Word2Vec:**

Word2vec is a two-layer neural net that processes text. Its input is a text corpus and its output are a set of vectors: feature vectors for words in that corpus. While Word2vec is not a deep neural network, it turns text into a numerical form that deep nets can understand [6]. This model is provided by Google and is trained on Google News data. This model has 300 dimensions and is trained on 3 million words from google news data. Team used skip-gram and negative sampling to build this model

### 5. One hot encoding

One hot encoding is a process by which categorical variables are converted into a form that could be provided to machine learning and deep learning algorithms to do a better job in prediction. It is a group of bits among which the legal combinations of values are only those with a single high (1) bit and all the others low (0). A similar implementation in which all bits are '1' except one '0' is sometimes called one-cold. [7]

### B. Information Extraction for Event Detection

Information extraction is associated with template-based extraction of event information from natural language text, which was a popular task of the Message Understanding Conferences (MUC) in the late eighties and nineties. MUC information extraction tasks started from a predefined set of templates, each containing specific information slots that encode event types relevant to a very specific subject domain – for instance, terrorism in Latin America – and used relatively straightforward pattern matching techniques to fill out these templates with specific instances of these events from a corpus of texts. Patterns in the form of a grammar or rules (e.g., in the form of regular expressions) were mapped on the text in order to identify the information [8]. Information extraction is used to get some information out of unstructured data. Written and spoken text, pictures, video and audio are all forms of unstructured data. It identifies information in texts by taking advantage of their linguistic organization. It is traditionally applied in situations where it is known in advance which kind of semantic information is to be extracted from a text. For instance, it might be necessary to identify what kind of events are expressed in a certain text and at what moment these events take place [8]. Among the most important tasks of information Extraction, we can mention the followings:
- Noun Phrase Co-Reference Resolution
- Named Entity Recognition
- Semantic Role Recognition
- Entity Relation Recognition
- Timex and Time Line Recognition

- **Named Entity Recognition**

Named entity recognition (NER) is one of the important tasks in information extraction, which involves the identification and classification of words or sequences of words denoting a concept or entity. Examples of named entities in general text are names of persons, locations, or organizations. Domain-specific named entities are those terms or phrases that denote concepts relevant to one particular domain. For example, protein and gene names are named entities that are of interest to the domain of molecular biology and medicine. The massive growth of textual information available in the literature and on the Web necessitates the automation of identification and management of named entities in text.

### 1. Named Entity Recognition Approaches

Named entity recognition activities began in the late 1990's with limited number of general categories such as names of persons, organizations, and locations. Early systems were based on the use of dictionaries and rules built by hand, and few used supervised machine learning techniques. The CoNLL-03 [9] provided valuable NE evaluation tasks for four languages: English, German, Spanish, and Dutch. With the extension of named entity recognition activities to new languages and domains, more entity categories are introduced which made the methods relying on manually built dictionaries and rules much more difficult to adopt, if at all feasible.

### 2. Common machine learning architecture for NER

Constructing a named entity recognition solution using a machine learning approach requires many computational steps including preprocessing, learning, classification, and post-processing.

RELATED WORKS

Detecting and disseminating food hazard information is a critical task that directly affects the public health. In spite of its importance, however, few systems are developed to automatically gather and analyze food hazard information.
Letia Ioan Alfred et al [10] proposes an agent-based solution to support small and medium companies willing to implement a Hazard Analysis Critical Control Point system in food supply chains. This approach generates two ontologies to differentiate the hazard effects. Hazards are defined as biological, chemical, or physical agents that are likely to cause illnesses or injuries to the consumer. Ming Yi [11] performed an analysis of the various food safety-strengthening requirement that might cause the food hazards in case of violation of parameter values. They proved that the implementation of HACCP principles can more effectively strengthen the food circulation safety and resolve many food safety issues arising in China's food circulation field, including: secondary contamination of food circulation, "three- no products" in the rural market, food safety issues in the circulation field caused by uncritical check on production process. Hu Boran [12] attempted to ensure the food safety by detecting the food hazard related parameters by integrating the Petrinet with the HACCP. This is analyzed and evaluated in the strawberry production process. This analysis work builds the prevention and control measures which is collected from the strawberry food products. In order to gain the cause of hazard according to HACCP correctly, HACCP and Petri net are combined and applied to study the food safety and look for the cause of the food hazard. Zhou Qiang et al [13] analyze the reason that led to the malfunction and inefficient status of food safety crisis management of China. It mainly focusing on problems of overlapping functions; overstaffing; stagnant information changing; inadequate legal protection; poor quality officers; obstructions of restriction in trade association; lacking of food security and social responsibility conscious; low participation of public and media between different departments. Hu Weizhong & Jin Han [14] explores factors influencing the efficacy of risk communication of food safety. The study shows consumer's risk perception of food safety changes rapidly according to the content of information. It seems that consumer's risk perception of food safety is unstable in that risk perception can be easily reversed by information opposite to consumer's previous attitude. Liang Meiyu et al [15] introduced the novel method that focus on detecting the semantic topic models from the food hazard events to improve the food safety system. It applies the CHCLDA method to establish the model for the news topics and reports, so as to realize the topic modeling in the semantic feature space. It can resolve the problems of high dimension and sparseness in the feature space and semantic relevance, and improve the time efficiency for LDA method to realize the semantic mapping of the feature space. Law Whisker TY et al [16] introduce an Advanced Rapid Alert System (ARAS) to effectively deliver food safety alerts in a timely manner with structured information to identify affected trades related to problematic food lots and prevent them from selling as soon as possible. The ARAS requires information integration from various government departments and public services through Web services and Service Oriented Architecture (SOA). Han Pengcheng et al [17] designed and implemented a cross-media information retrieval system based on the area of food safety emergencies. The system collects Internet information using topic crawler, establishes data index on cross-media information and makes fast retrieval by sort labeling. Li Hui et al [18] constructed the system of total merit index of food safety warning and the technique of quantitatively calculating different levels is given. Consequently, it contributes to scientific grounds for an objective, comprehensive and deep analysis in online information of food safety sector. Jihong Zhang [19] proposed a framework of food safety alert system and put forward a new model for food safety management in China based on statistical analysis of food quality information on the Internet. Li Fang et al [20] analyze event detection in the perspective of multiple data sources, combining news reports and micro blogs. Detect events from news, combining micro blogs to do event monitoring and early warning. Jang Kyoungrok et al [21] introduce the preliminary work to build such system. The final system aims to detect and extract food hazard event from the live data shared on the Web. They defined information template for food hazard event. Then used it to extract informative keywords from the website of Ministry of Food and Drug Safety, the governmental organization responsible for ensuring food safety in Korea.

PROPOSED SYSTEM

This section focuses on the description of our proposed approach. The details will be outlined through several sections including system description, system architecture, and data set.

- **System Description**

The system aims to collect, extract and treat the data and food hazard information from social media such as Facebook, technical reports and websites that are concerned by food security. The data can be either texts, or texts and images. The system consists of a set of processes implying an intense use of data mining algorithms and concepts in addition to machine learning and deep learning. The idea is to use this data as input for the system that seeks to convert texts into Named Entities for food hazard such as: {*Type of food hazard, Place where it were founded, the Organization or Persons that found it, Quantity of the hazard, and Date of the event*}. The extracted information is then used to fill a Food hazard Template with the new entities.

- **System Architecture**

The overall system will be composed of two main components: Text Analysis and Information extraction. These two processes will be at the heart of the system, which has as input a collection of Arabic texts and as output a set of entities. The text analysis process receive data from the local dataset 'Food Hazard Database' previously collected from the Arabic social media, websites and technical reports pertaining to the food domain. Once the data is analyzed and processed by the first process, a food hazard template will be filled using the data received as output from the previous process. The results of the template will be converted into a visualization format to display it as a warning system or used to feed any related decision support system. Figure 6 is showing the architecture of whole system with focus on data transmission between different steps. Since our main concern is to deal with Arabic datasets, the focus of this project is on text pre-processing as well as named entity recognition. Once these components are implemented, a machine-learning algorithm can then be applied to complete the data analysis phase. The next sections will give a brief overview of each step with special focus on Text Analysis. Particularly the Text preprocessing as well as with the Named Entity Recognition process will take most of our attention. Figure 1 is showing the Framework architecture of our proposed system.

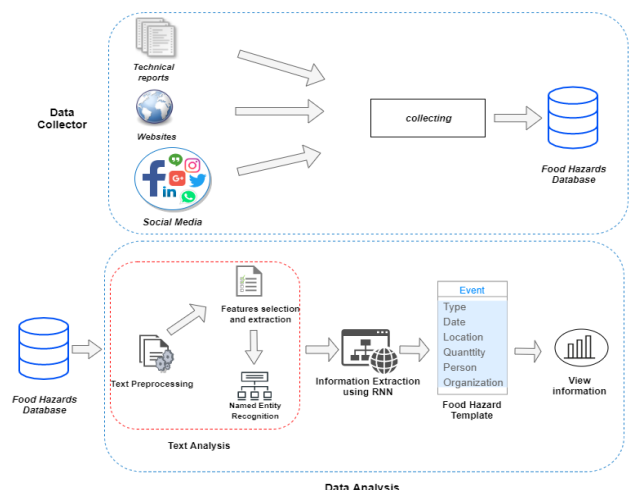

Figure 1. System Architecture

1. **Dataset**

Seemingly, all dataset related to food hazard data are in English language. According to our knowledge, the studies that used the Arabic content were almost non-existent in this field. The fact that there is no ready-made Arabic data set to be used by the system will push us to prepare our in-house collection of Arabic texts in the field of the food hazard. Consequently, our data set will be collected from technical reports, websites, and social media posts. Our primary data was collected by three different ways from:

- *Technical reports*: the Hygiene and Security Commission affiliated to commune of Setif.
- *Websites*: Health Ministry, Trade Ministry, Consumer Protection Organization.
- *Social media*: From Facebook: Posts that are related to food & food hazard.

## 2. Text Analysis Phase

The aim of this phase is to prepare the text's content to be ready for task concerned by this research study.

**Text Pre processing**

This phase will start by getting the Arabic texts from the dataset. Therefore, it needs to go through many steps of text mining pre-processing and natural language processing (NLP) operations. Thus, we will describe a practical example of each step as follows:

***Word tokenization:*** Tokenization is one common task in Natural Language Processing. "Tokens" are usually individual words (at least in languages like Arabic) and "tokenization" is taking a text or set of text and breaking it up into its individual words. These tokens are then used as the input for other types of analysis or tasks, like parsing (automatically tagging the syntactic relationship between words). For example: " تمكنت اليوم السبت لجنة النظافة والأمن " will be transformed into "تمكنت" , "اليوم" , "السبت" , "لجنة" , "النظافة" , "و" , "الأمن" .

***Stemming:*** Because of documents are using different forms of a word, such as تمكن ,تمكنت and تمكنوا, this operation aim to return the Arabic keywords that are related to food hazard to their canonical roots.

For example: مصابين and إصابة will be transformed to أصاب, which means in English language "infected".

All these steps of preparing text content can be done using any text mining or NLP tools like *KNIME* Text processing, *RapidMiner* Text Extension, *Apache OpenNLP* library, Natural Language Toolkit *IBM SPSS*, *FARASSA* or Arabic Linguistic Tool *ALP* …etc.

***Content transformation***: this step will constitute the text's content as a vector of weighted features. The weight of features will be counted based on the number of how many times the words was appeared in the text's content. Many techniques can be used for this aim like N-grams, which is a sub of words or character from the original text. For example, when we take a sentence from the texts such as " تمكنت لجنة النظافة والأمن " we can identify and form the following N-Grams:
- Three unigrams: ("تمكنت" , " لجنة " , " النظافة " , " والأمن").
- Two bigrams ("تمكنت لجنة"," النظافة والأمن")
- One trigram ("تمكنت لجنة النظافة والأمن").
Each n-gram among them will be counted as one attribute or feature.

**Features selection and extraction:**

This step will start after of the end of the text pre-processing operation. Features selection techniques are generally used to improve the performance of machine learning algorithm. More specifically this operation can be very useful in terms of dimensionality reduction because the learning algorithm will deal only with most important and representative features. Therefore, the feature extraction step will reduce the value of feature space when the value in data input is very large then to be replaced with smallest representation set. Thus, there are many methods and techniques to achieve this kind of task such as PCA, LSA, Chi-Square or LDA.

## 3. Named Entity Recognition Phase

After completing the Text Analysis Phase steps, a machine-learning algorithm can then be used to build and train an intelligent model from the data that was preprocessed. The algorithms are divided into three main categories based on the of learning model: supervised, semi-supervised and unsupervised learning. More precisely, detecting named entities in free unstructured text from data received from previous steps is not a trivial task. There have been many approaches to build rule-based models to detect Named Entities; however, these models fail to scale and work on limited vocabulary. The modern-day solution to this problem is to use Deep Neural Networks to classify sequences of text in order to tag a sentence into a sequence of tags. Depending on our task and our needs, the used tags will be the followings: {*O, PERSON, LOCATION, ORGANIZATION, QUANTITY, EVENT and DATE*}.
- *O*: means short for outside, this is the class of the words that mark no entities. These words represent 40% ality of the text and it includes stop-words words like "هذا", "على", "تمكنت", "في", "غير", "من", "أكثر", "و"…
- *PERSON (PERS)*: This is the class of person's names. Example: "مفتشوا النظافة".
- *LOCATION (LOC)*: This the class of locations. It can be a city, a country, a town or a continent. Example: سطيف.
- *ORGANIZATION (ORG)*: This is the class of organizations, like: "لجنة النقاوة العمومية ", "لجنة النظافة والأمن".
- *QUANTITY (QUANT)*: this is the class of quantities. Example: "28 كلغ", "70 كلغ".
- *EVENT (EVT)*: this is the class of the main key word in the title of the food hazard event like: " اللحوم الحمراء و "مادة المرقاز" , "البيضاء"
- *DATE (DTE)*: this is the class of dates. Example: "26 ماي 2019", "19 ماي 2019".

The most famous scheme for representing these classes is the *IOB* representation where in this scheme classes are appended with either a B- or an I- which indicates that this class is a "Beginning" or an "Inside" class of a Named Entity, for example the input sentence:
"حجز أكثر من قنطار من اللحوم الحمراء في سطيف" would be tagged as:
"O /O /O /B-QUANT /O /B-EVENT /I-EVENT /O /B-LOC".

## 4. Information Extraction System

In this phase, the system will fill the food hazard template, which contain food hazard event's information according to entities obtained from the Named Entity phase.

• **System Components**

The whole system is composed of three major components including dataset (Food Hazard Database), Text Analysis and Information Extraction. Each component is directly depending in terms of inputs to the output of its predecessor. Each component can consist of one or more processes that must be executed sequentially. As shown in Figure 4, the text analysis phase will obtain the input data from the food hazard database. The text's content is preprocessed based on text preprocessing operation tokenization, stemming and text's content transformation. After that, the feature selection and extraction will be applied in order to reduce the dimensionality and to decrease sparsity. This last is used by the machine learning algorithm to predict and to classify the selected words into theirs final categories (location, quantity, date…etc). The last step will be the visualization of the system's results which means the translation of the new detected knowledge to a new format that can be understood and useful by others people. In this case, a warning system will be a good choice for this system. Figure 2 is showing the details of our system components.

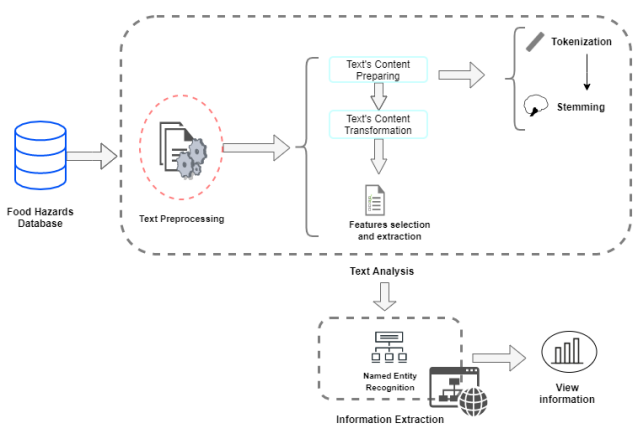

Figure 2. System Components.

- **Evaluation metrics**

The last work phase will be the evaluation phase that will focus on the assessment of the system's performances. This phase will use some known evaluation metrics that are generally applied in the different types of application such as recommender systems (RS), information retrieval (IR) systems and Information extraction (IE) systems. In this phase, we will measure these metrics based on confusion matrix that will show if the keywords in text's content matches to the categories of the named entities or not. We will utilize several metrics such as F1, accuracy, precision and recall [56].

CONCLUSION

This work represents a case study and aims to develop and propose a new system based on deep learning to extract events related to food hazards from social media. To achieve research objectives, we proposed an information extraction system that will be based on a recurrent neural networks and LSTMs. For future work, we would study further many related problems and test our system on large datasets for better performance. Another perspective is to deal with food hazard images. We would like to collect images related to food hazards during the phase of events detection and try to identify their contents using a deep learning algorithm like Convolutional Neural Network.